\documentclass[useAMS,usenatbib]{mn2e}
\usepackage{graphicx}

\title[Near-IR studies of novae V2491 Cyg and V597 Pup]{Near Infrared Observations of the novae V2491 Cygni and V597 Puppis} 
\author[Naik, Banerjee $\&$ Ashok]{Sachindra Naik\thanks{snaik@prl.res.in}, D. P. K. Banerjee\thanks{orion@prl.res.in}, \& N. M. Ashok\thanks{ashok@prl.res.in}
\\
Physical Research Laboratory, Navrangapura, Ahmedabad - 380009, Gujarat, India}
\begin{document}

\date{Accepted for publication in MNRAS}

\maketitle


\begin{abstract}
We present results obtained from near-infrared $JHK$ spectroscopic
observations of  novae V2491~Cyg and V597~Pup in the early declining 
phases of their 2007 and 2008 outbursts respectively. In both objects, 
the spectra displayed  emission lines of HI, OI, HeI and NI. In V597 Pup, 
the HeI lines were found to strengthen rapidly with time.  Based on the 
observed spectral characteristics, both objects are classified as He/N 
novae. We have investigated the possibility of V2491 Cyg being a recurrent 
nova  as has been suggested. By studying the temporal evolution of the line 
widths in  V2491~Cyg  it appears unlikely that the binary companion is  
a giant star with heavy wind as in recurrent novae of the RS Oph type. 
Significant deviations from that of recombination case B conditions are
observed in the  strengths of the HI lines. This indicates that the  HI 
lines, in both novae,  are optically thick during the span of our 
observations. The slope of the continuum spectra in both cases was 
found to have a  $\lambda^{-(3-3.5)}$ dependence which deviates from 
a Rayleigh-Jeans spectral distribution. Both novae were detected in 
the post-outburst super-soft X-ray phase; V2491 Cyg being very bright 
in X-rays has  been the target of several observations. We discuss and 
correlate our infrared observations with the observed X-ray properties 
of these novae.  
\end{abstract}


\section{Introduction}
We present here near-infrared studies of two classical novae V2491 Cygni
and V597 Pup. The observations cover the early declining phase in both 
these novae. Both objects have been detected in X-rays but V2491 Cyg  
has specifically been the subject of considerable current interest. 
It is one of the rare novae to have been detected in X-rays even before 
its recorded optical outburst (Ibarra \& Kuulkers, 2008). Apart from 
that, it has also shown very prominent and sustained super soft X-ray 
emission in recent times as  described later. Our infrared observations, 
in conjunction with observations from other wavelength regimes like the 
X-ray and optical, should give a more comprehensive understanding of 
these interesting objects.

The characteristics of a nova outburst depend on three principal 
parameters viz. the white dwarf mass, the core temperature of the 
white dwarf and the rate of mass accretion from the binary companion. 
The accreted mass which eventually ignites the explosion is ejected 
back into space at high velocities of the order of a few hundreds 
to thousands of kilometers per second. A white dwarf accreting mass  
below the critical rate undergoes a thermonuclear runaway in the 
hydrogen burning shell at the bottom of the accreted layer and becomes 
a classical nova when sufficient amount of mass has been accumulated.  
The brightness of a classical novae begins to decline once mass ejection 
subsides after the ejection of most of the envelope. As the remnant of 
the envelope collapses back onto the white dwarf, residual hydrogen 
burning occurs in the shell radiating most of the energy in the UV to 
X-ray band. The photosphere of the white dwarf shrinks which increases 
the effective temperature and the nova appears as a hot blackbody-like 
object. The temperature of the blackbody-like object (post nova) is 
predicted to be about 3$\times$10$^5$ K for classical novae for a few 
years after optical maximum (Prialnik 1986). The post-outburst nova 
with this high temperature photosphere is a source of  super-soft 
X-rays which lasts for a duration depending on the leftover 
envelope mass, chemical composition of the envelope, mass of the white 
dwarf etc. (Kato 1997; Hachisu \& Kato 2006). Seven such super-soft 
Galactic novae appear to have been observed with ROSAT and Swift till 
2006 June 30 (Orio et al. 2001; Ness et al. 2007). Some of the novae 
also show hard X-ray emission, the origin of which is different from 
the emission from the white dwarf photosphere. X-ray emission from the 
white dwarf is not expected to be observed immediately after the nova 
outburst. X-ray detections in some of the novae during this phase is 
understood to  originate from some other processes such as shocks within 
the expanding shell or from a collision of the expanding shell and the 
atmosphere of the companion as seen in RS Opiuchi (Bode et al. 2006). 
Therefore, it is rather interesting to understand and relate the 
mechanism of classical nova outburst and the ensuing super-soft 
X-ray emission phase.

V2491~Cyg was discovered on 2008 April 10.728 UT at a visual magnitude
of 7.7 (Nakano et al. 2008). Low resolution spectra of the source 
obtained on April 11.72 UT showed prominent broad Balmer emission lines 
indicating the object to be a nova in its early phase of outburst (Ayani 
\& Matsumoto 2008). Subsequent infrared observations on April 12.56 UT 
confirmed the object as a nova with an unusual spectrum with extremely 
broad lines of complex profiles (Lynch et al. 2008). The presence of 
the helium line at 1.083 $\mu$m and strong emission features of NI and 
NII in the near-infrared spectrum suggested the object to be a 
helium-nitrogen (He/N)nova. Balmer lines from H$\alpha$ to H$\delta$ 
in emission were seen in the optical spectrum taken on April 11.99 UT 
and April 13.95 UT (Tomov et al. 2008). The FWHM of the H$\alpha$ line
decreased from $\sim$4800 km s$^{-1}$ to $\sim$4600 km s$^{-1}$ within 
about 2 days. Another near-IR observation on April 17.6 UT (Rudy et al. 
2008) found that the brightness of the nova had already declined by a 
factor of three  to  photometric magnitudes of J=7.7, H=7.8 and K=7.4; 
this was accompanied by  an  increase in the strengths of the emission 
lines. The OI lines at 0.8446 and 1.1287 $\micron$, which are fluorescently 
excited by Lyman $\beta$, were found to be the strongest emission 
features in the infrared spectrum. The relative strength of these lines 
imply  a reddening of E(B-V) =  0.43 (Rudy et al. 2008). Broad HI emission 
lines and prominent HeI and OI lines were detected in the near-infrared 
spectra of the nova on April 18 and 20 (Ashok et al. 2008). They have 
also pointed out the possibilities of the presence of NI lines at 
1.1762 and 1.2462 $\micron$s. A periodicity of 0.0958 day with  0.03-0.05 
magnitude amplitude variations was suggested from the optical photometric 
observations (Baklanov et al. 2008).

Nova V597~Pup was discovered on 14 November 2007 at a visual magnitude
of 7 (Pereira et al. 2007). Low resolution spectrum of V597~Pup on 
November 14.77 UT in the 410-670 nm range showed the presence of broad 
Balmer lines of Fe II with P-Cyg features along with the blue-shifted 
absorption lines of Fe II (49, 74), OI  and Si II (Naito \& Tokimasa 
2007). The expansion velocities on the same day were estimated to be 
1800, 1700, and 1800 km s$^{-1}$, derived from the FWHM of the Balmer 
H$\alpha$, H$\beta$ and H$\gamma$ emission lines respectively. The 
near-infrared spectrum (in the 0.8-2.5 $\micron$ range) of the nova 
on November 30, 2007 showed that the He~I and O~I lines were very 
strong and the Br~$\gamma$ line showed a double-peaked line profile, 
without any evidence of He~II lines and thermal emission from dust; 
a value of E(B-V) = 0.3 was estimated (Rudy et al. 2007). Similar IR 
observations on 2008 January 7 showed the presence of strong He~II 
lines and weak coronal lines of [S~VIII], [Si~VI] and [Ca~VIII] in 
the  spectrum which was accompanied by the appearance of  super-soft 
X-rays from the nova (Ness et al. 2008a).

Both  V2491~Cyg and V597~Pup were detected in the post-outburst, super-soft 
X-ray phase with various X-ray observatories. In case of V597~Pup, the 
X-ray emission was detected on 2008 January 8 and 17 from the Swift 
satellite (Ness et al. 2008a).  Most of the photons detected were below 
0.7 keV indicating the onset of the super-soft X-ray phase. These were 
the only observations  reporting the X-ray detection in V597~Pup. However, 
V2491~Cyg was much brighter in X-ray during the super-soft X-ray phase 
and hence observed with Swift, Suzaku, and XMM-Newton observatories. 
Pre-outburst observations with Swift showed the presence of a variable 
soft X-ray source at the position of V2491~Cyg  even three months before 
the outburst. This makes the object specially interesting since it is 
only the second nova after V2487 Oph (Hernanz \& Sala 2002) to be detected 
in X-rays even before eruption (Ibarra et al. 2008).   Nearly continuous 
monitoring of the nova from April 17 to May 14 with Swift found that the 
brightness of the source increased continuously since April 27, reaching 
$\sim$200 count ks$^{-1}$ on May 14 with the significant increase below 
1.5 keV (Page et al. 2008). The search for any periodicity in X-ray on 
the longest Swift observations on May 8 and 10, to corroborate the 
suggested 0.1 day orbital period (Baklanov al. 2008),  yielded a negative 
result. In addition to the Swift monitoring, Suzaku observation of the 
nova on May 9 showed the presence of strong low energy emission  from 
K-shell N, O, and Ne lines in the spectrum (Osborne et al. 2008). Swift 
observations on May 16 showed a very strongly increasing count rate which 
touched $\sim$5 counts s$^{-1}$ on May 18. XMM-Newton observations of 
the nova, on 2008 May 20, detected several O and N absorption lines 
along with the broad Ne emission lines, suggesting the nova to be an 
ONe nova (Ness et al. 2008b). 

\begin{figure*}
\vskip 7.5 cm
\includegraphics{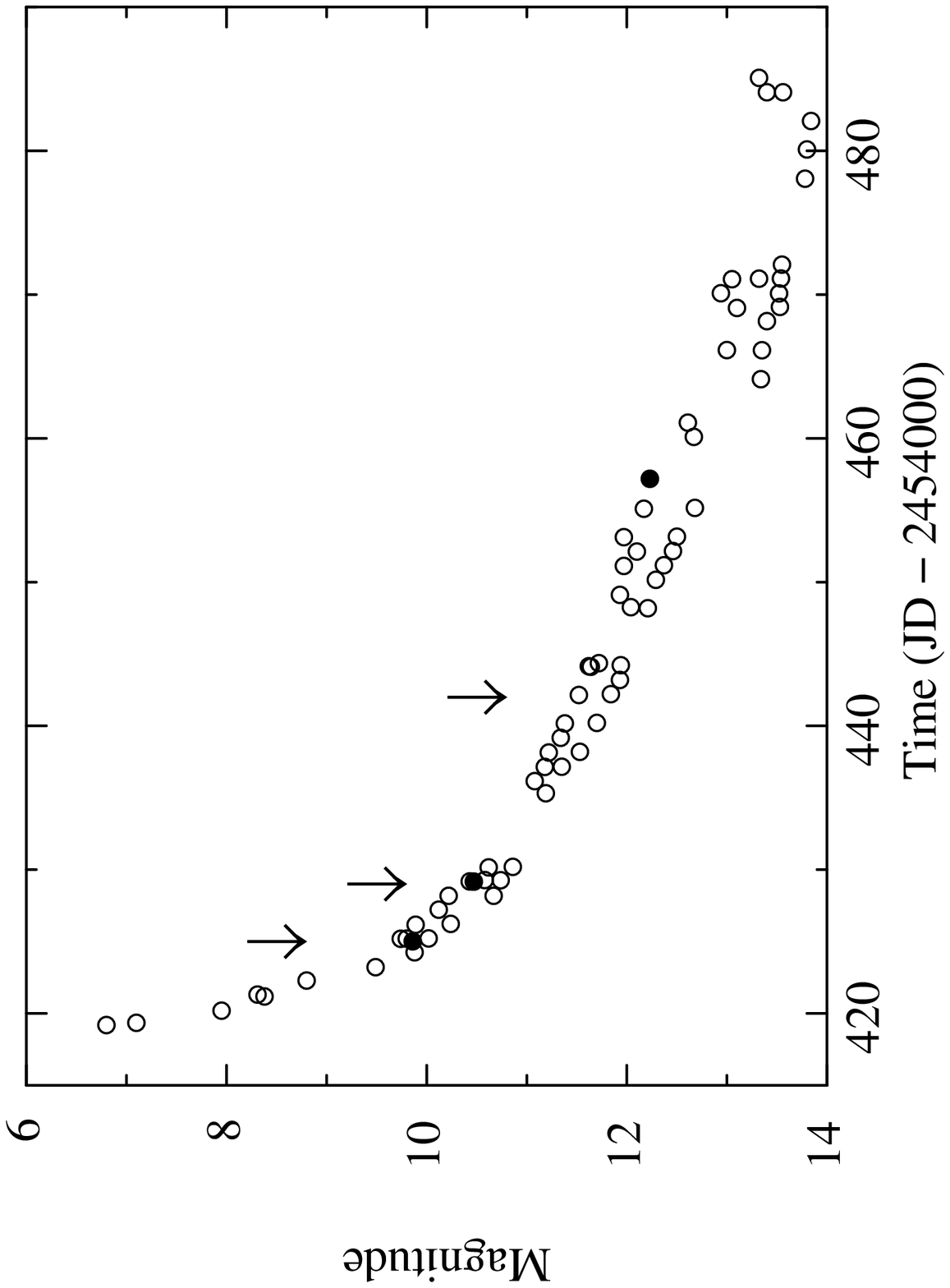}
\includegraphics{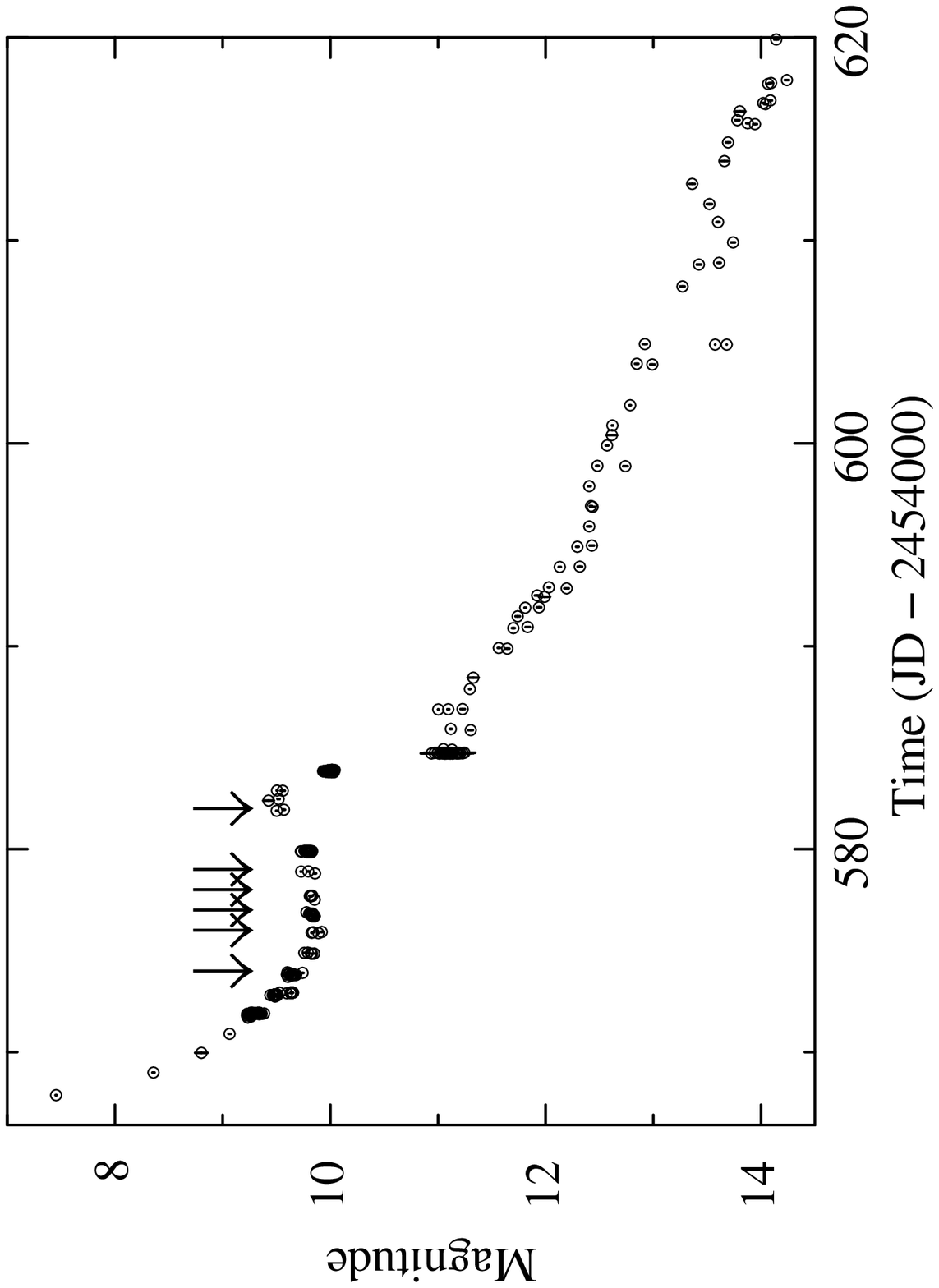}
\caption{The light curves of V597~Pup (left panel; V band 
data from AAVSO shown in filled circles while the V band data from  
AFOEV is shown in open circles) and V2491~Cyg (right panel; using  
V band data from the AAVSO). The arrow marks in both the panels 
show the days of our near-infrared observations.}
\label{fig1}
\end{figure*}

\section{Observations and Analysis}

The optical light curves of  V2491 Cyg and V597~Pup are shown in 
Figure~\ref{fig1}. While V597 Pup showed a gradually declining 
light curve,  V2491~Cyg showed  an unusual second rebrightening with 
$\sim$2 magnitude amplitude peaking after about 17 days of the main 
outburst. The nature of this second mini-outburst is rather similar 
to those observed in V1493 Aql and V2362 Cyg (Venturini et al. 2004; 
Kimeswenger et al. 2008). From the light curves we estimate $t_2$ to 
be 2.5 and 5 days for V597 Pup and V2491 Cyg respectively which 
indicate they are fast novae. The $t_2$ values yields  distance 
estimates of 10.3 and 11.4 Kpc, for V597 Pup and V2491 Cyg 
respectively, from the MMRD relation of della Valle \& Livio (1995). 
The near-infrared observations of the novae with the Mt Abu telescope 
are shown with arrow marks on the light curves in Figure~\ref{fig1}.  
The log of the observations along with the signal-to-noise ratio 
(S/N) of the spectra of V597~Pup and V2491~Cyg, is given in Table~1.

Near-Infrared spectroscopic observations of   V597~Pup and V2491~Cyg
were carried out in $JHK$ bands at the Mt Abu 1.2m telescope in the 
early declining phases of 2007 November  and 2008 April outbursts 
respectively. The  spectra were obtained at a resolution of $\sim$1000 
using a Near-Infrared Imager/Spectrometer with a 256$\times$256 HgCdTe 
NICMOS3 array. Spectral calibration was done using the OH sky lines 
that register with the stellar spectra. The spectra of the standard 
stars SAO~198195, SAO~174400 (for V597~Pup) and SAO~67663, SAO~68744, 
SAO~88071 (for V2491~Cyg) were taken at similar airmass as  the 
corresponding targets to ensure  the ratioing process (nova spectrum 
divided by the standard star spectrum) removes the telluric lines 
reliably. To eliminate artificially generated emission lines in the 
ratioed spectrum – created due to the H~I absorption lines in the 
spectrum of the standard star – the hydrogen absorption lines in 
the spectra of the latter were removed by interpolation before 
ratioing. The ratioed spectra were then multiplied by a blackbody 
curve corresponding to the standard star's effective temperature 
to yield the final spectra. 

Photometric observation of V2491~Cyg could only be done on one night
viz. 2008 April 22 in photometric sky conditions using the NICMOS3 array 
in the imaging mode. Several frames in each of the three J, H and K bands 
were obtained in four dithered positions, offset by typically 30 arcsec 
from each other. The sky frames were generated by median combining the 
dithered frames and  subtracted from the nova frames - aperture photometry 
was subsequently done using IRAF. A nearby star in the field, with known  
2MASS magnitudes, was used as standard star for photometric calibration.

\begin{table}
\centering
\caption{Log of the  near-infrared spectroscopic observations of 
V597~Pup and V2491~Cyg. The date of outbursts have been assumed to be their 
discovery date viz. 2007 November 14.23 UT and 2008 April 10.728 UT, 
respectively.} 
\begin{tabular}{lccccc}
\hline
\hline
Date of   &Days            &Sp. &Int.  &S/N   &Standard\\
Obs.      &since           &band     &time &    &star\\
          &outburst        &         &(s)        &     & \\
\hline
\hline
\multicolumn{5}{|c|}{2007 November outburst of V597~Pup}\\
\hline
Nov. 21  &7    &J     &180   &18  &SAO~198195\\
         &     &H     &120   &41  &  \\
         &     &K     &180   &28  &  \\
Nov. 25  &11   &J     &120   &17  &SAO~174400\\
         &     &H     &180   &35  &  \\
         &     &K     &180   &13  &  \\
Dec. 08  &24   &J     &360   &13 &SAO~174400\\
         &     &H     &360   &24 & \\
         &     &K     &360   &10 & \\
\hline
\multicolumn{5}{|c|}{2008 April outburst of V2491~Cyg}\\
\hline
April 18   &8      &J    &120  &28  & SAO~67663\\
           &       &H    &80   &29  &          \\
           &       &K    &120  &26  &          \\
April 20   &10     &J    &180  &26  & SAO~68744\\
           &       &H    &180  &42  &          \\
           &       &K    &240  &16  &          \\
April 21   &11     &J    &180  &29  & SAO~68744\\
           &       &H    &180  &52  &          \\
           &       &K    &240  &21  &          \\
April 22$^a$ &12   &J    &150  &24  & SAO~88071\\
           &       &H    &80   &44  &          \\
           &       &K    &120  &15  &          \\
April 23   &13     &J    &150  &27  & SAO~88071\\
           &	   &H	 &120  &37  &\\
	   &	   &K	 &120  &15  &\\
April 26   &16     &J    &120  &19  & SAO~88071\\
           &       &H    &120  &31  &          \\
           &       &K    &150  &15  &          \\
\hline
\hline
\end{tabular}
$^a$ : Photometric observation  of V2491~Cyg was also carried out
on 22 April yielding J, H and K magnitudes of 8.22$\pm$0.03, 
8.32$\pm$0.03 and 7.96$\pm$0.05 respectively.  \label{table1}
\end{table}

\begin{figure*}
\centering
\includegraphics[bb= 76 317 371 686, width=2.2in,height=3.2in,clip]{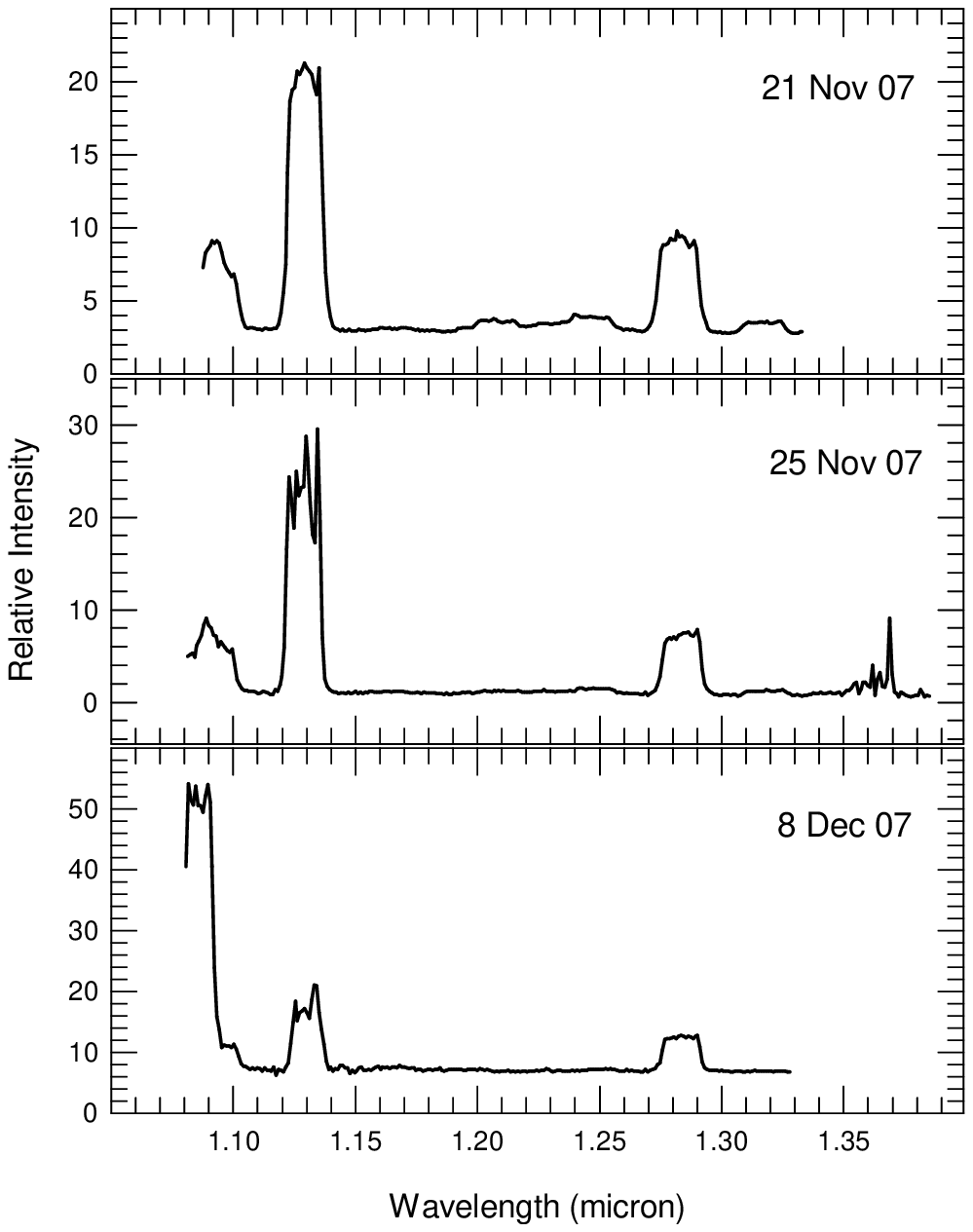}
\includegraphics[bb= 119 419 414 711, width=2.2in,height=3.2in,clip]{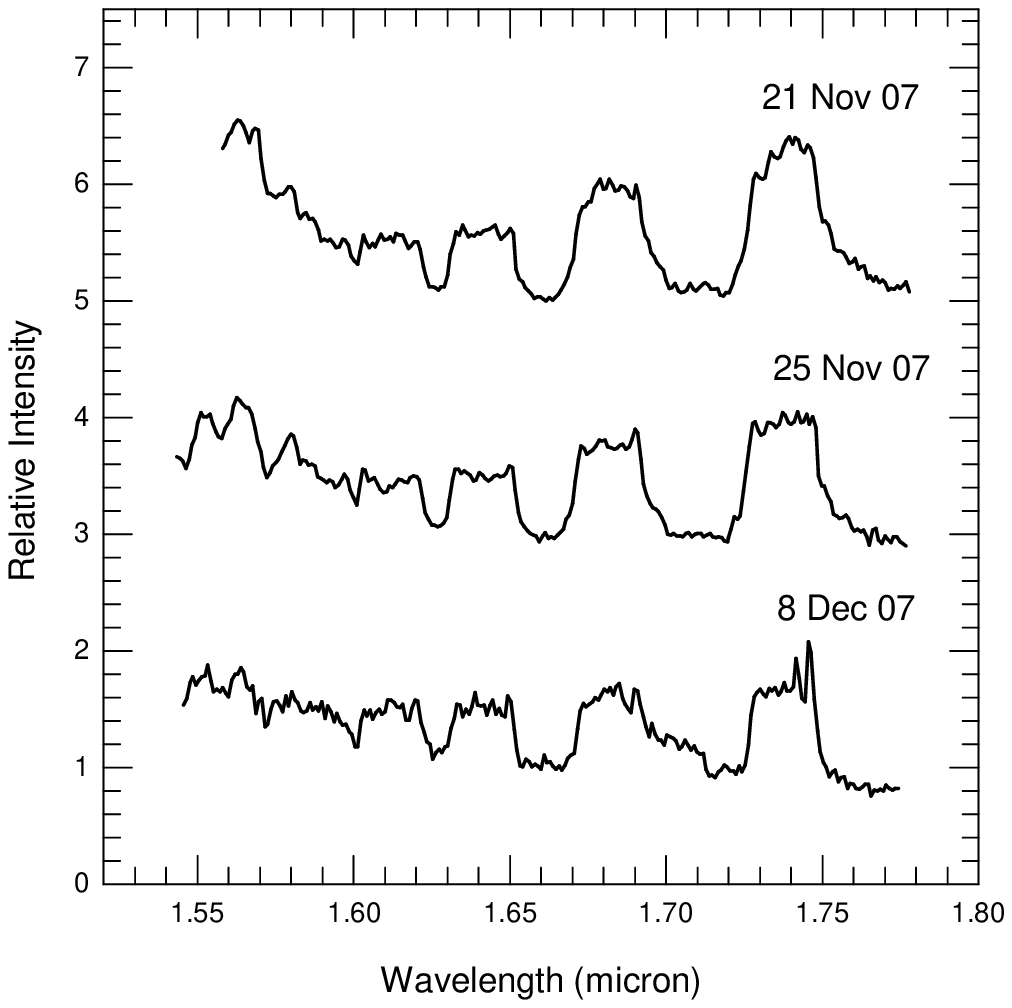}
\includegraphics[bb= 81 277 382 574, width=2.2in,height=3.2in,clip]{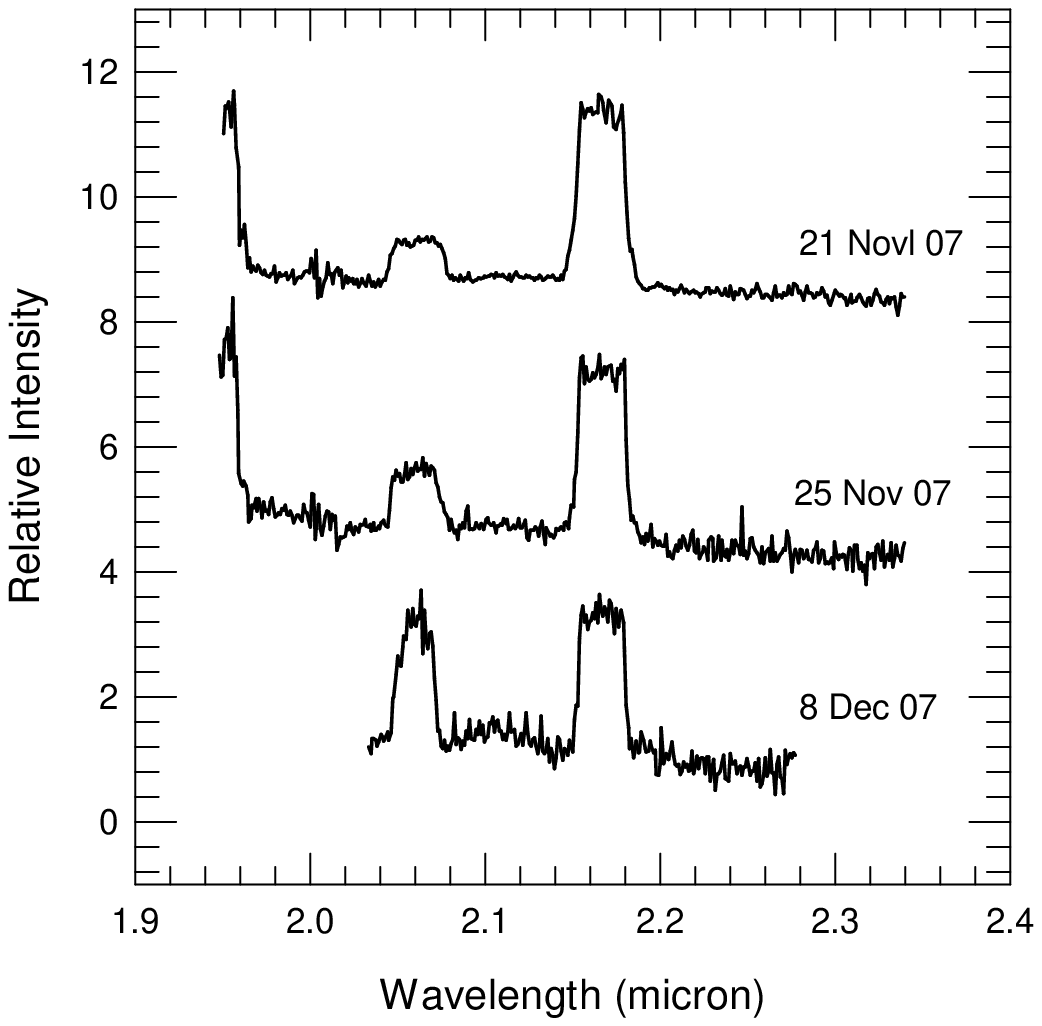}
\caption{The $JHK$  spectra of V597~Pup at different epochs are 
shown in the left, middle and right panels respectively. The $JHK$ spectra 
have been offset from each other for clarity. The $J$ band spectra have 
been plotted in different sub-panels to accommodate the strong changes 
in the HeI 1.0830 \micron~ line strength between epochs.}
\label{fig2}
\end{figure*}

\begin{figure*}
\centering
\includegraphics[bb= 105 273 409 680, width=2.2in,height=3.8in,clip]{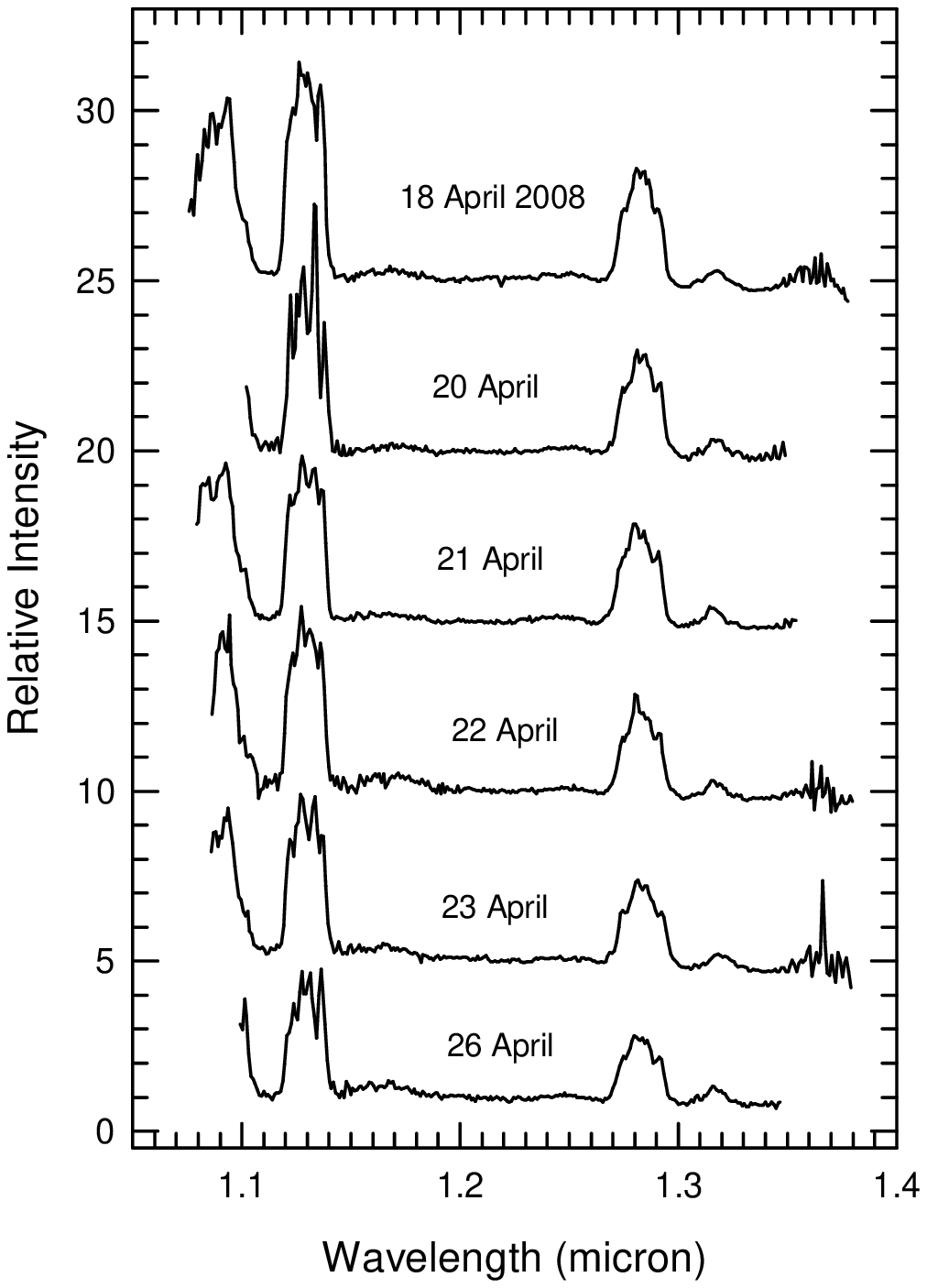}
\includegraphics[bb= 100 241 403 665, width=2.2in,height=3.8in,clip]{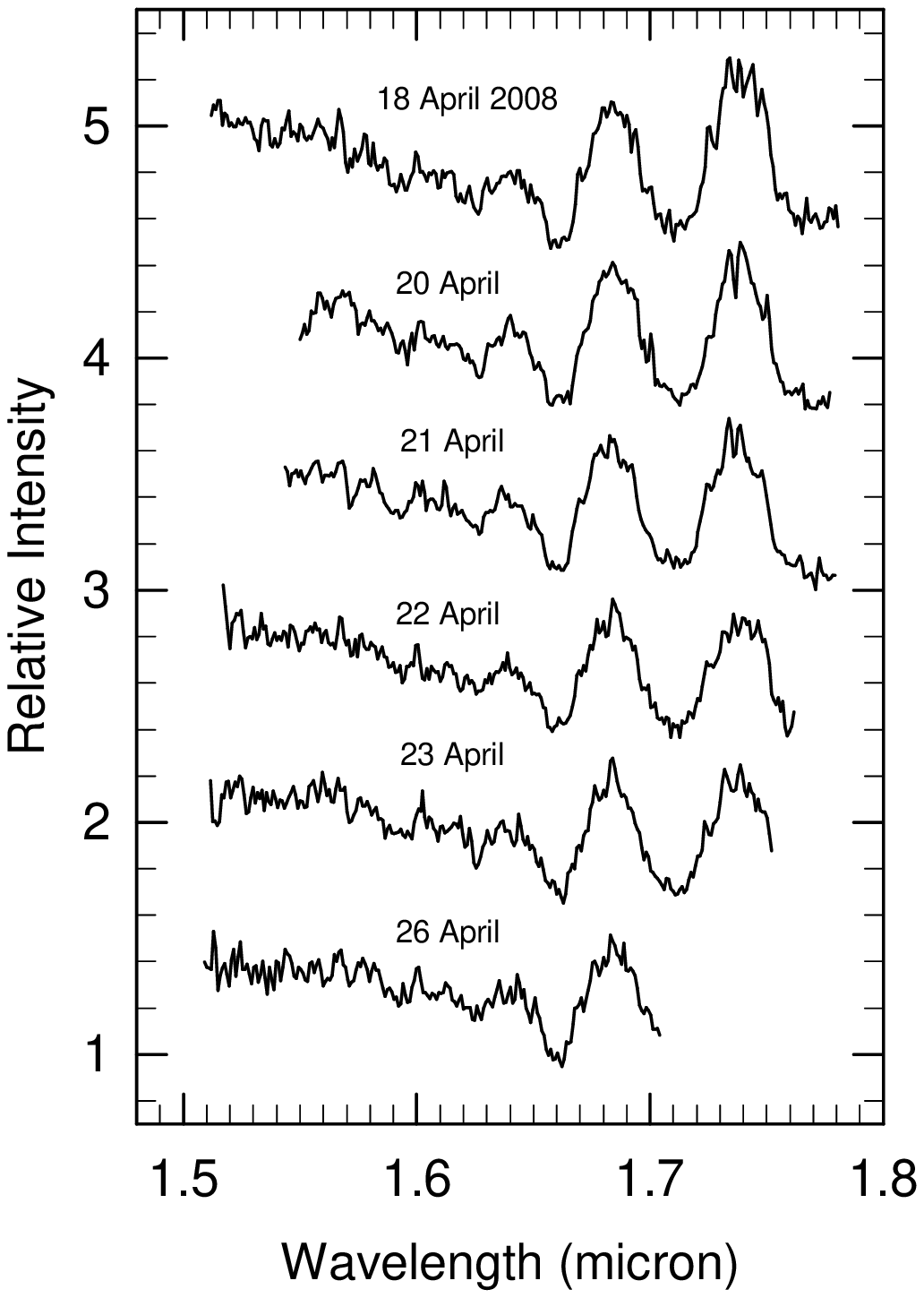}
\includegraphics[bb= 141 219 445 635, width=2.2in,height=3.8in,clip]{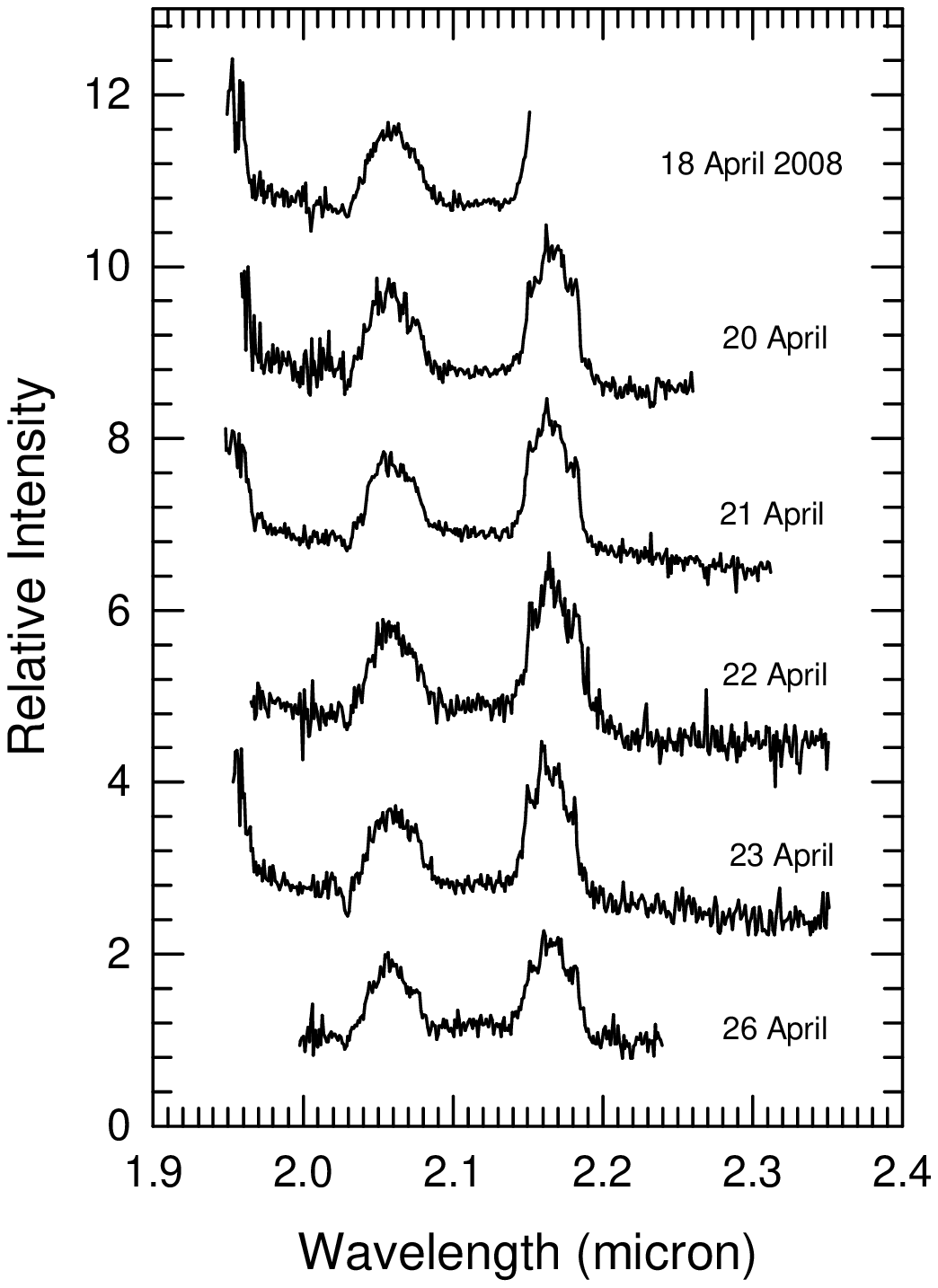}
\caption{The $JHK$ band spectra of V2491~Cyg at different epochs are shown
in the left, middle  and right panels respectively. The spectra have 
been offset from each other for clarity.}
\label{fig3}
\end{figure*}

\section{Result and Interpretation}
\subsection{General characteristics of the $JHK$ spectra of V2491~Cyg 
and V597~Pup }
The $JHK$ band spectra of V597~Pup and V2491~Cyg at different epochs 
are presented  in  Figure~\ref{fig2} \& \ref{fig3} respectively. The 
prominent features detected in the spectra of both the novae are the 
He~I  1.0830 \micron, Pa$\gamma$  1.0938 \micron, O~I  1.1287 \micron, 
Pa$\beta$  1.2818 \micron, He~I  2.0581 \micron~ and hydrogen Brackett 
series lines. A few relatively weaker NI lines are seen in the spectra 
of both objects. In V2491 Cyg, a line-like feature  at $\sim$1.36 
\micron~ is seen. Though it is susceptible to an error in identification  
since it is at the edge of the spectral window,  we feel reasonably 
certain  that this line  is due to NI 1.3602-1.3624 \micron. It may also 
be mentioned that in the $JHK$ bands, there are certain regions with low 
atmospheric transmission e.g. the spectral region around the O~I 1.1287 
\micron~line and also that between 2 to 2.05 \micron. In regions of
such difficult atmospheric transmission, the process of ratioing the
source spectrum with the standard star spectrum can produce some errors 
or artifacts in the final extracted spectrum. Thus the structures that 
are seen in the peak of the OI 1.1287 \micron~ line should be treated with 
some caution. The reliability of these structures may be cross-checked
by comparing them with the profiles of other O~I line profiles (e.g. the
0.8446 \micron~ line) in case optical spectra of the object are available on
similar dates. The details of the line identification are given in Table~2. 
No significant evolution is seen in the lines in the case of V2491~Cyg 
during the 9 day span of the observations. However, in the case of V597~Pup, 
it is found that the He~I line at 1.0830 \micron\ strengthens dramatically 
(by a factor of about 7 within 17 days) by 2007 December 8 (Figure~\ref{fig2}). 
On December 8, the HeI 1.7 \micron~ line  is also prominently seen along 
with  weaker lines of HeI 2.1120-2.1132 \micron. The  rapid strengthening 
of the HeI lines, also noted by  Rudy et al. (2008),  indicates a quick 
evolution to higher excitation conditions in the ejected envelope. In 
both the novae, no coronal line features were  detected till the last 
day of our campaign. This is consistent with the non-detection of both 
the novae in super-soft X-ray phases during the entirety of our 
observations.

Based on several optical and near-IR spectra taken after the outburst, 
V2491 Cyg has been classified as a He/N nova (Tomov et al. 2008; Lynch 
et al. 2008; Helton et al. 2008). From our spectra (Fig 3) we would 
agree with this classification  based on the early appearance of 
prominent HeI lines  and also the presence of NI lines. Based on the 
optical spectrum obtained on November 14.77, Naito and Tokimasa (2008) 
suggest V597 Pup to be of the FeII type. However we note that V597 Pup 
has a very similar near-IR spectrum compared to V2491 Cyg, which  indicates it 
is also a He/N nova. In further support of this He/N classification, we 
note that neither V597 Pup nor V2491 Cyg   show the typical IR spectrum 
expected of FeII novae (or equivalently CO novae) early after outburst. 
One of the principal IR signatures of such novae is the presence of 
prominent CI lines - especially in the J band but also in the K band 
(these CI lines  are not seen here). Typical spectra of CO novae in 
the near-IR can be seen in the cases of  V2274~Cygni \& V1419~Aql 
(Rudy et al. 2003) and V1280 Sco  (Chesneau et al. 2008, Das et al. 
2008).  

\subsubsection{V2491 Cyg: a possible recurrent nova?}
\begin{figure}
\centering
\includegraphics[bb=150 184 440 543, width=3.0in,height=3.7in,clip]{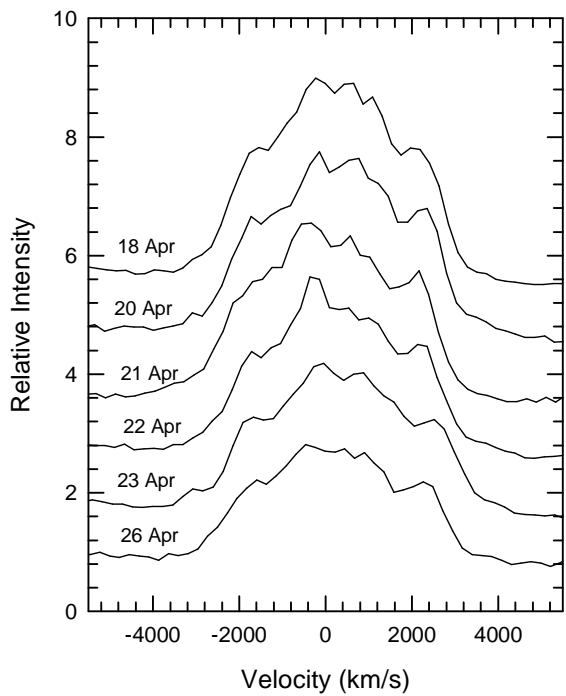}
\caption{Evolution of Pa$\beta$ line throughout the near-infrared 
observations of V2491~Cyg. There appear to be marginal differences 
in the line profiles during the  observations probably due to the 
density inhomogeneities in the ejected envelope. The overall width 
of the line did not change appreciably.}
\label{fig4}
\end{figure}

There has been some discussion  whether V2491 Cyg could be a recurrent
nova (RN) based on the nature of its optical spectrum. Tomov et al. (2008)
pointed out the similarities between the  15 April and 17 April 2008 
optical spectra of V2491~Cyg with those  of  recurrent novae V394~CrA 
and U~Sco recorded in earlier outbursts. Attention has also been drawn
in VSNET (Japan)\footnote{http://ooruri.kusastro.kyoto-u.ac.jp/pipermail/vsnet-newvar/2008-April/000139.html} alerts to the similarity of the optical 
spectrum of V2491~Cyg on 20 April 2008\footnote{Spectra taken from Bisei 
Observatory; http://www1.harenet.ne.jp/\~{}aikow/etc/v2491\_cyg\_20080420.gif} 
(10 days after the outburst) to that of the recurrent nova V394~CrA on 3 
August 1987 (about 5 days after the outburst; Sekiguchi et al. 1989). The 
similarity is indeed striking  though there are some differences: e.g., the 
peak of H$\alpha$ line  is a sharp peak in the case of V2491~Cyg whereas it is 
flat-topped in the case of V394~CrA. A thorough search in archival patrol 
plates should be able to give a more definitive answer to whether V2491 
Cyg has had earlier outbursts - snapshot images of the object as in Palomar, 
Supercosmos plates etc. do not show any changes in the brightness of the  
progenitor. The recurrent nova hypothesis can be tested, to some extent, 
from the IR data present here. If it is indeed a recurrent nova of the 
RS Oph type, then considerable decrease  in the width of the line 
profiles is expected with time. In RS Oph type systems the secondary 
is a high mass losing red giant and the ejected nova material is severely 
decelerated as it plows through the wind of the companion. This 
leads to rapid temporal changes in the expansion velocity of the 
ejecta - an effect that is well documented for the 2006 outburst 
of RS Oph in Das et al. (2006). In order to quantify the changes 
in the line profile widths, which effectively measure the expansion 
velocity, the evolution of the prominent emission lines in V2491~Cyg 
was investigated. Figure~\ref{fig4} shows the Pa$\beta$ line profile 
for all six days of our observations of V2491~Cyg.  The Pa$\beta$ 
line is chosen because, among the HI lines, it is observed with a 
high  S/N on all 6 days of observation.  There appear to be some 
differences in the structure of the profile which is probably due 
to the inhomogeneities in the ejected envelope. Visual inspection 
shows that the overall width of the emission lines (Figure~\ref{fig4}) 
does not change appreciably during the observations. The FWHM of 
the Pa$\beta$ line (with one sigma uncertainty) is estimated to be 
3950$\pm$65, 3975$\pm$48, 4085$\pm$34, 3971$\pm$87, 4053$\pm$110, 
3920$\pm$112 km s$^{-1}$ on 2008 April 18, 20, 21, 22, 23, and 26 
observations of V2491~Cyg respectively. This indicates very marginal 
change in the line width during the span of our observations. This 
is unlike the findings in the recurrent nova RS~Oph in which the 
width of the emission lines started decreasing within only a few days 
of the nova outburst (Das et al. 2006).  However, this is not the case 
in V2491~Cyg. Thus even if V2491 Cyg is  a RN, it certainly is not of 
the RS Oph category and we can rule out the possibility of a red giant 
nature for the binary companion of V2491~Cyg. Further, the discovery 
of the variability in V and R  spectral band of V2491~Cyg with a period 
of $\sim$0.1 day (Baklanov et al. 2008) suggested the nova to be a  
binary system with orbital period of $\sim$0.1 day. This value of binary 
period is extremely small compared to that of the recurrent novae binaries 
with giant binary companions.

\begin{table}
\begin{center}
\caption{List of observed lines in the $JHK$ spectra}
\begin{tabular}{l l l | l | l l}
\hline
\hline
\multicolumn{2}{|c|}{V2491~Cyg}  &\multicolumn{2}{|c|}{V597~Pup}\\
\hline
Wavelength        &Species   & Wavelength  & Species \\
(${\rm{\mu}}$m)   &          & (${\rm{\mu}}$m)   &         \\
\hline 
\hline  
1.0830   &He I$^a$           &1.0830   & He I$^a$           \\
1.0938   &Pa $\gamma$$^a$    &1.0938   & Pa $\gamma$$^a$    \\
1.1287   &O~I                &1.1287   & O I  \\
1.1625, 1.1651   &N~I       &1.1625, 1.1651   &N I  \\
	 &              &1.1969   &He I     \\
         &              &1.2074   &N I      \\
1.2461, 1.2470   &N~I       &1.2461, 1.2470   &N I  \\  
1.2818   &Pa $\beta$    &1.2818   &Pa $\beta$ \\
1.3164   &O~I           &1.3164   & O I     \\
1.3602, 1.3624   &N~I      &         &  \\ 
1.6109   &Br 13         &1.6109   & Br 13   \\
1.6407   &Br 12         &1.6407   & Br 12    \\
1.6806   &Br 11         &1.6806   & Br 11    \\
	 &              &1.7002   &He I      \\
1.7362   &Br 10         &1.7362   & Br 10    \\
1.9446   &Br 8          &1.9446   & Br 8     \\
2.0581   &He I          &2.0581   &He I      \\
         &              &2.1120, 2.1132   &He I \\
2.1655   &Br $\gamma$   &2.1655   &Br $\gamma$ \\
\hline
\hline
\end{tabular} 
\\
\end{center}
$^a$ : He~I~1.0830 \micron\ \& Pa~$\gamma$ lines are blended.\\
\label{line-info}
\end{table}

\subsection{The shape and evolution of the Continuum}

\begin{figure}
\centering
\includegraphics[bb= 90 255 450 482, width=3.3in,height=3.0in,clip]{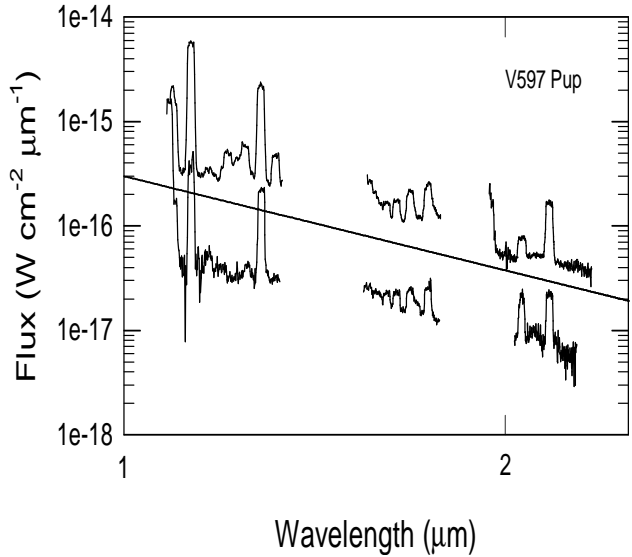}
\caption{The composite $JHK$ band spectra of V597~Pup are shown for 
observations of 2007 November 21 (top spectrum) and December 8 (bottom 
spectrum).  The slope of the continuum  ($F_\lambda~~\propto~~\lambda^\alpha$)
is compared with a line of slope $\alpha$ = -3.}
\label{fig5}
\end{figure}

\begin{figure}
\centering
\includegraphics[bb= 112 265 477 487, width=3.3in,height=3.0in,clip]{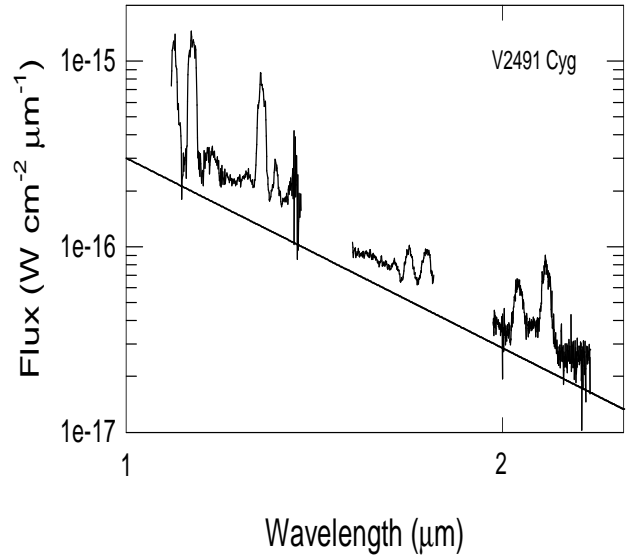}
\caption{The composite $JHK$ band spectrum of V2491~Cyg of 2008 April 22 
observations is shown.   The slope of the continuum  ($F_\lambda~~\propto~~
\lambda^\alpha$) is compared with a line of slope $\alpha$ = -3.4.}
\label{fig6}
\end{figure}

The shape of the continuum of both  novae was also studied for any
significant changes with time. The flux calibrated $JHK$ band 
spectra for V597~Pup and V2491~Cyg are plotted in Figure~\ref{fig5} \& 
\ref{fig6} respectively. As we did not have any photometric observations 
of V597~Pup, our spectra for 2007 November 21 \& 2007 December 8 were 
flux calibrated by taking appropriate values of magnitudes from the 
nova light curve obtained from the KANATA 
telescope\footnote{http://kanatatmp.blogspot.com}. Though, there is 
a possibility of some error in estimating the exact values of the 
continuum flux while using values taken from a figure, the effect is 
not significant when comparing the shape of the continuum. Figure~\ref{fig5}
shows the flux calibrated spectra of V597~Pup for November 21  
and December 8 observations, along with a line ($F_\lambda ~\propto 
~\lambda^\alpha$) drawn in between for a slope of $\alpha$ = $-$3.
For V2491~Cyg, we calibrated the spectrum of 2008 April 22 using  
our photometric $JHK$ magnitudes of 8.22$\pm$0.03, 8.32$\pm$0.03 
and 7.96$\pm$0.05 respectively obtained on the same day. This flux 
calibrated spectrum  is shown in Figure~\ref{fig6} along with a 
line drawn for a slope of $-$3.4.

The characteristics and evolution of a nova's continuum spectrum are 
not too well understood. At the outburst maximum, during the fireball 
expansion phase, the  pseudo-photosphere is well reproduced by a 
black-body continuum of a A or F spectral type star. Subsequently the 
continuum deviates from a blackbody and can evolve into a free-free 
continuum. This is particularly well seen in the case of Nova 
Cygni~1975 in the 1-20 \micron~ spectral region which showed a 
spectral change from  a Rayleigh-Jeans spectrum to that of thermal 
bremsstrahlung emission within a short period of  2 to 3 days (Ennis 
et al 1977). The near-infrared continua of V597~Pup and V2491~Cyg,
however, do not show any significant changes in the shapes of the 
continua. The slopes of the continuum spectra in these novae were 
found to be $\sim-3.0$ and $\sim-3.4$. Though the deviation is not
much from the Rayleigh-Jeans type spectrum with slope $-4$, it might
be slowly changing towards the thermal bremsstrahlung emission.
This slow change in the slope of the continuum spectrum is also
evident in case of other novae e.g. V4643 Sagittarii where it changed 
from a slope of about $-3$ to $-2$ within about three months (Ashok 
et al. 2006). However in the case of  another nova viz. V4633~Sagittarii 
(Lynch et al. 2001) the change in the slope of the continuum was found 
to be in the opposite direction i.e. the slope changed from $-2$ to
$-2.7$ in observations taken 525 and 850 days after outburst . It is, 
therefore, necessary to follow the nova outburst more regularly to 
have a better spectral coverage to understand the behavior of the 
nova continua.

\subsection{Recombination Case B Analysis}
\begin{figure}
\centering
\includegraphics[bb= 149 191 445 548, width=3.2in,height=3.2in,angle=-90,clip]{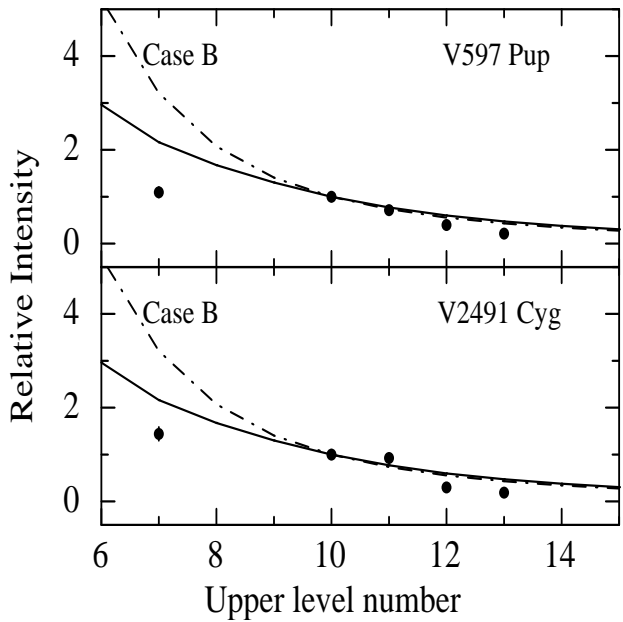}
\caption{Optical depth effects seen in the hydrogen Brackett 
lines in V2491~Cyg and V597~Pup. The abscissa is the upper level number
of Brackett series line transition. The line intensities are relative 
to that of Br~10 which is normalized to unity. The errorbars in the line
intensities ($\sim$10\% for all the lines other than Br13 where it is about 
20\%) are small compared to the size of the symbol. The case~B values for 
the line strength are also shown for temperature $T=10^4 K$ and electron 
density of $n_e = 10^{12} cm^{-3}$ (shown as dotted lines) and for 
temperature $T=10^4 K$ and electron density of $n_e = 10^{10} cm^{-3}$ 
(solid line).} 
\label{caseb}
\end{figure}

The presence of HI lines in the near-infrared spectra of V597~Pup and 
V2491~Cyg permit a recombination analysis to be done. In both the novae, 
Pa$\gamma$ at 1.0938 \micron~  is strongly blended with the HeI  1.0830 
\micron~  line thereby making it difficult to assess the formers strength
correctly. Hence in our analysis, among the HI lines, we consider only 
the intensities of Pa$\beta$ at 1.2818 \micron~ and the Br series lines.
Among the Br series lines we restrict ourselves to Br$\gamma$, Br10, Br11, 
Br12, and Br13 lines. Higher Br series lines like Br14, Br15 etc. although 
present are too severely blended with each other to allow accurate 
extraction of their equivalent widths and hence intensities. We have 
done the recombination analysis for the April 22 and November 25 spectra 
of V2491 Cyg and V597 Pup respectively. In Figure~\ref{caseb}, we have 
plotted the observed strength of the Brackett series lines along with
the predicted values in two recombination case~B conditions. The 
errors in the estimated values of the strength of the lines relative
to that of the Br10 line are $\sim$10\% for all the lines other than 
the Br13, where it is about 20\%. The errorbars in Figure~\ref{caseb} 
are smaller than the size of the symbols and hence not clearly visible. 
The case~B line intensities are taken from Storey \& Hummer (1995) for
a representative temperature of $T = 10^4 K$ and electron densities of 
$n_e = 10^{10} cm^{-3}$ and  $n_e = 10^{12} cm^{-3}$.  High electron 
densities are considered because the ejecta material is expected to 
be dense in the early stages after the outburst.  Figure~\ref{caseb}
shows that the observed line intensities clearly deviate from the
optically thin case~B values. In particular, Br$\gamma$, which is 
expected to be significantly stronger than Br10 or Br11, is instead 
observed to be of similar strength.  We have not included Pa$\beta$ 
in Figure 7 but  mention that the observed ratio of Pa$\beta$ to
Br$\gamma$ is observed to be 9.4 and 9.2 in V2491 Cyg and V597 Pup 
respectively. Such observed ratios are again significantly different 
from the expected values of 5.2 and 6.5 for the above Case B conditions. 
The above analysis indicates that the ejecta of both the novae appear 
to be optically thick to the Brackett and Paschen lines during the 
span of our observations. Such optical depth effects during the early 
stages after outburst can be expected and have been observed in other 
novae too e.g. V4643 Sagittarii (Ashok et al. 2006) and V4633~Sagittarii 
(Lynch et al. 2001).

\section{Discussion}

While our IR observations do not throw direct light on the interesting 
X-ray behavior of V2491 Cyg, there are  some correlated aspects which 
could be discussed. But before proceeding to do so, it is appropriate to 
briefly discuss  the super soft X-ray phase in novae. 

The discovery of X-ray emission from nova GQ~Mus, 463 days after the 
outburst (Ogelman et al. 1984), indicates that  novae in  post-outburst 
should go through an extended phase of residual hydrogen burning and 
appear as a super-soft X-ray source (SSS). The duration of the post-outburst 
super-soft X-ray phase depends on the nuclear burning of the residual 
mass on the surface of the white dwarf and also on the mass of the white
dwarf. Massive white dwarfs show super-soft X-ray phases for short duration.
This is because the stronger gravity in massive white dwarfs finishes the
hydrogen burning in the residual envelope faster compared to that of less
massive white dwarfs. The theoretical estimation of the duration of the 
residual hydrogen burning (super-soft X-ray phase) ranges from 5.6 years 
to 2$\times$10$^4$ years for white dwarf masses from 1.35 to 0.6 
M$_{\odot}$ (Gehrz et al. 1998). However, observations showed a much 
shorter duration (a few hundred days) of the post-nova super-soft X-ray 
phase (Pietsch et al. 2005) implying that either the mass of the white 
dwarfs is much higher than the commonly assumed mass or the X-ray turn-off 
depends on some other parameters other than the white dwarf mass. Along 
with V597~Pup and V2491~Cyg, the super-soft X-ray emission has been 
detected, so it would appear, in fewer than a dozen novae (Orio et al. 
2001; Drake et al. 2003; Ness et al. 2007 and references therein; Ness 
et al. 2008a, 2008b). 

While the SSS phase is believed to arise from the hot photosphere of 
the white-dwarf, we investigate whether a fraction of the observed X-rays 
from V2491 Cyg could arise from an alternative mechanism viz. a shock in 
the ejecta. A plausible reason to expect a shock is the secondary outburst 
seen in V2491 Cyg. If this mini-outburst is due to a significant second 
mass-loss phase in which the ejected matter moves at a higher velocity 
than the primary ejecta, then this  matter  could catch up to and collide 
with the earlier expanding envelope leading to the formation of a shock. 
The hot shocked gas could then contribute to the observed X-rays. Hence 
it is necessary to check whether there is any significant increase in the
velocity profiles during the second rebrightening.  However, from Figure 4, 
we do not see any evidence for any enhancement in velocity for the matter 
ejected around the peak of the mini-outburst ($\sim$ 26 April). Therefore 
it is unlikely that a shock could have formed due to the mini-outburst.   
The optical light curve of V2491~Cyg resembles  that of  two other novae, 
V1493~Aql and V2362~Cyg, in terms of the presence of a secondary peak. The 
secondary peaks in V2491~Cyg and V1493~Aql (Venturini et al. 2004)  were 
observed about 16 and 48 days after the primary outburst respectively, 
where as in case of V2362~Cyg it was after $\sim$240 days (Kimeswenger 
et al. 2008). Among these three novae, the SSS phase has been detected 
in V2491~Cyg (present work) and V2362~Cyg (Hernanz et al. 2007). Because 
of only three novae showing the secondary peak in the light curve, out 
of which only two are detected in X-ray, it is premature to say whether 
the secondary peak is linked to the detected X-ray emission from the 
novae.

The beginning of the SSS phase (X-ray turn-on) i.e. the residual 
nuclear burning in various novae is generally detected after more than 
a hundred days of outburst. Some of the examples are, 250 days after the 
outburst in V1974~Cyg (Krautter et al. 1996), 200 days in Nova LMC 1995 
(Orio et al. 2003), 180 and 222 days  in V382~Vel (Orio et al. 2002), 
about 250 days  in V1494~Aql (Drake et al. 2003), about 180 days in 
V4743~Sgr (Ness et al. 2003) and about 175 days in V574~Pup (Ness et 
al. 2007). However, in case of V597~Pup and V2491~Cyg, the super-soft 
X-ray state was detected about 55 and 30 days after outburst respectively, 
which is  rather early compared to the other novae discussed above. As 
per the model of Hachisu \& Kato (2006) the relatively early onset of the 
SSS phase in these novae suggests  the white dwarfs in these systems are 
very massive and close to the Chandrasekhar limit.

\section*{Acknowledgments}
The research work at Physical Research Laboratory  is funded by the
Department of Space, Government of India. We express our thanks to 
the referee Dr. Ray W. Russell for valuable comments which improved
our paper. We thank Vishal Joshi for helping with some of the observations. 
We acknowledge with thanks the variable star observations from the AAVSO 
International Database contributed by observers worldwide and used in 
this research. This research has made use of the AFOEV database, operated 
at CDS, France.

\end{document}